\documentclass[conference,10pt]{IEEEtran}
\IEEEoverridecommandlockouts
\usepackage{cite}
\usepackage{amsmath,amssymb,amsfonts}
\usepackage{algorithmic}
\usepackage{graphicx}
\usepackage{textcomp}
\usepackage{braket}
\def\BibTeX{{\rm B\kern-.05em{\sc i\kern-.025em b}\kern-.08em
    T\kern-.1667em\lower.7ex\hbox{E}\kern-.125emX}}
    
\UseRawInputEncoding

\begin{document}

\title{The Exact Entropy Formula of the Ideal Gas and its
Information-Theoretic Interpretation
\\
}

\author{Arnaldo Spalvieri \\
Dipartimento di Elettronica, Informazione e Bioingegneria \\
Politecnico di Milano\\
arnaldo.spalvieri@polimi.it \\
ORCID 0000-0002-8336-7996
}

\maketitle

\begin{abstract}

The paper analyzes the entropy of a system composed by
non-interacting and indistinguishable particles whose quantum
state numbers are modelled as independent and identically
distributed classical random variables. The crucial observation is
that, under this assumption, whichever is the number of particles
that constitute the system, the occupancy numbers of system's
quantum (micro)states are multinomially distributed. This
observation leads to an entropy formula for the physical system,
which is nothing else than the entropy formula of the multinomial
distribution, for which we claim novelty, in the sense that it is
proposed here for the first time that the entropy of the
multinomial distribution is the entropy of the physical system.
The entropy formula of the multinomial distribution unveils yet
unexplored connections between information theory and statistical
mechanics, among which we mention the connection between
conditional entropy of the random microstate given the random
occupancy numbers and the Boltzmann-Planck entropy $\log(W)$ and
between these two and the Gibbs correction term $\log(N!)$,
thermalization and communication-theoretic preparation of a
thermal state, accessible information of the thermal state and
physical entropy of the thermalized system. A noticeable specific
result that descends from our approach is the exact quantum
correction to the textbook Sackur-Tetrode formula for the entropy
of an ideal gas at the thermal equilibrium in a container.
\vspace{0.5cm}

{\em Keywords:}\hspace{0.2cm}{\bf Entropy, Mutual Information,
Holevo Information, Equiprobability of Microstates, Ideal Gas,
Sackur-Tetrode Formula, Szilard Engines.}

\end{abstract}

\section{Introduction}

The concept of physical entropy is controversial since the times
of Boltzmann and Gibbs. The following passage of an interview with
Shannon can be found in \cite{tribus}:

{\em My greatest concern was what to call it. I thought of calling
it 'information,' but the word was overly used, so I decided to
call it 'uncertainty.' When I discussed it with John von Neumann,
he had a better idea. Von Neumann told me, ''You should call it
entropy, for two reasons. In the first place your uncertainty
function has been used in statistical mechanics under that name,
so it already has a name. In the second place, and more important,
no one really knows what entropy really is, so in a debate you
will always have the advantage.''} \newline During the years, the
number of different interpretations and definitions of entropy has
grown, a recent collection of heterogeneous ''entropies'' being
reported in \cite{obsafra}.

In this fragmented and ambiguous context, one of the most debated
points is the relationship between information and physical
entropy. The controversial about the role of information in
physics plunge its roots in the famous thought experiments of
Maxwell and Szilard, and is still today object of discussion. From
the one side, Landauer claims in \cite{land2} that

{\em Information is a physical entity}. \newline From the other
side, Maroney writes in his thesis \cite{mt} that

{\em The Szilard Engine is unsuccessful as a paradigm of the
information-entropy link,} \newline and Maroney and Timpson
reiterate the same concept in \cite{maroneyrecent}:

{\em rejecting the claims that information is physical provides a
better basis for understanding the fertile relationship between
information theory and physics.}

Whatever are the different opinions and arguments, it is a matter
of fact that, starting from the concept of entropy that
information theory and statistical mechanics share, many authors
built in the past bridges between them, a prominent example being
\cite{neritut}. Making a comprehensive review of the bibliography
that links information theory and statistical mechanics is far
from the objectives of this author. In this introductive section,
we mention only few papers whose scope we find to be closer to the
scope of the present paper. Other bibliography relevant to
specific points touched in the paper will be cited in the body of
the paper where the specific point is discussed. With reference to
the connections between classical thermodynamics and classical
information theory, a special place must be deserved to the
pioneering work of Jaynes \cite{maxent}, after which the principle
of maximum entropy has been universally recognized and accepted.
The connections between information theoretic inequalities and the
irreversibility of certain thermodynamical processes have also
been deeply investigated in the context of information theory, see
\cite{merhav} and Chapter 4 of \cite{cover}. Paper
\cite{spalvieri} has recently shown that, under certain assumption
on particles' quantum states, information-theoretic typicality
applies to system's microstates. One of the works that more
contributed to the strengthen the link between the two disciplines
is that of Landauer \cite{landauer}, that showed the equivalence
between heat and logical information. This major result is today
proved experimentally \cite{landdemonstration} and it is widely
accepted that (quantum) thermodynamics can be treated by (quantum)
information-theoretic tools, see for instance the tutorial papers
\cite{plenio,trends,itthermoreview,parrondo}.

In this paper, we try to contribute to the link between
information theory and statistical mechanics by proposing a model
in which physical entropy, intended in the standard thermodynamic
sense of a state variable, for instance, the entropy of a gas in a
container at the thermal equilibrium, matches the
information-theoretic definitions of classical mutual information
and of quantum accessible information.
Following our proposed approach we find an entropy formula that,
in the case of a gas at the thermal equilibrium in a container,
reduces to the Sackur-Tetrode formula when quantum effects are
neglected while, when quantum effects are taken into account,
unlike the Sackur-Tetrode formula, that can become negative at low
temperature, our entropy formula guarantees non-negative values of
entropy.  Our proposal is based on a radical novelty: the two
standard assumptions of non-interacting particles and of
equiprobability of microstates in the microcanonical ensemble are
replaced by the assumption that the quantum states of system's
particles are independent and identically distributed (i.i.d.)
random variables. We hasten to point out that we don't claim that
our new entropy formula always describes the entropy of the real
system, because there are certain cases where the i.i.d.
assumption could not hold for the real system. What we claim is
that our formula is the exact entropy formula whenever the i.i.d.
assumption holds.

The outline of the paper and specific claims of novelty follow. In
Section II we present the model of the system. Section III, which
is the core of the paper, defines the random occupancy macrostate
of the system as the vector of the occupancy numbers of the
quantum states, showing that these occupancy numbers are
multinomially distributed and that system's microstates are
conditionally equiprobable given the occupancy macrostate. These
two findings seem to be new. Section IV analyzes system's entropy
and its relationship with the mutual information between
microstate and macrostate that can be extracted from a measurement
made on a system prepared in an occupancy macrostate. All these
results, that are expressed in Sections III and IV  by the
language of classical random variables and classical information
theory, are presented in Section V by the formalism of quantum
mechanics and quantum information theory. Section VI analyzes the
concrete case of the thermal state of a monoatomic ideal gas in a
container. In this case, our exact entropy formula imports exact
quantum corrections in the standard Sackur-Tetrode entropy
formula, which are believed to be new. In Section VII we summarize
the main points of the paper and briefly sketch future work,
including the analysis of the Szilard engine.

\section{System model}

Let a statistical ensemble of systems made by $N$ particles of the
same species be represented by $\{p_{\bar{\cal
L}}(\bar{l}),{\bar{\cal L}}\}$, where the calligraphic character
denotes random variables, $p_{\cal X}$ denotes the probability
distribution of the discrete random variable (or vector) ${\cal
X}$, $\{{\cal X}\}$ denotes its support set and ${\bar{\cal L}}$
is a vector of $N$ random $D$-tuples of numbers, where the $i$-th
$D$-tuple represents by $D$ quantum numbers the quantum state of
the $i$-th particle. Hereafter the random vector ${\bar{\cal L}}$
is the energy eigenstate of the system that comes out from the
quantum measurement that projects the system onto its energy
eigenbasis. We look at this eigenstate as at the result of a
random experiment (the quantum measurement) performed on the
random ensemble $\{p_{\bar{\cal L}}(\bar{l}),{\bar{\cal L}}\}$.
The support set of ${\bar{\cal L}}$ is spanned by the vectors
$\{\bar{l}\}$ that appear in the argument of the probability
distribution. The $i$-th element $l_i$ of $\bar{l}$ is the
$D$-tuple of the quantum numbers that span the $D$-dimensional
energy eigenstate of the $i$-th particle,
\[\bar{l}=(l_{1},l_{2}, \cdots l_N), \ \ l_i=
(l_{i,1},l_{i,2}, \cdots l_{i,D}).\] In quantum statistical
mechanics, the quantum state of the system is called {\em
microstate} \cite{kardar}, so, in the language of statistical
mechanics, the statistical ensemble $\{p_{\bar{\cal L}}(\bar{l}),
\bar{\cal L}\}$ would be the {\em random} microstate, meaning the
set of microstates where the system can randomly collapse after
the already mentioned quantum measurement. Note that we do not
assume that the probability distribution $p_{\bar{\cal L}}$ is
uniform, so the ensemble $\{p_{\bar{\cal L}}(\bar{l}),{\bar{\cal
L}}\}$ is not the microcanonical ensemble, rather it is the
canonical ensemble. We anticipate that, in the following, we will
use the probability distribution $p_{\bar{\cal L}}$ to define the
entropy of the system. This is not a novelty. The definition of
entropy and of other thermodynamic quantities based on the
probability of the results of a quantum measurement made on the
system has been already proposed in \cite{plesch}. We will return
on this point in section IV. In the following we will omit the
adjective {\em random} when randomness is clear from the context,
for instance, we will write ''the probability distribution of
microstate(s)'', or ''entropy of microstate(s)''.

Suppose that an experimenter, that, in the following, we will call
Bob, wants to ascertain by a quantum measurement the quantum state
of the system. Bob has neither interacted in the past with the
system nor with the environment from which the system is separated
before his measurement. He is not informed about the past
interactions between the system and the environment and between
the particles of the system. More importantly, he is not
interested at all in these past interactions. These premises are
coherent with the goal of the paper, that is, proposing a model
for a variable of state, hence a variable that is independent of
how the system reached the state of interest. Given these
premises, the only reasonable approach of Bob to the measurement
is to assume that the elements ${\cal L}_1, {\cal L}_2, \cdots
{\cal L}_N,$ of $\bar{\cal L}$ are independent and identically
distributed (i.i.d.) random $D$-tuples:
\begin{equation}
p_{{\bar{\cal{L}}}}({\bar{l}})=\prod_{i=1}^N
p_{{\cal{L}}}({l}_i).\label{indep}
\end{equation}
Remarkably, the i.i.d. assumption makes it possible to present our
results in the next two sections without using the
quantum-mechanical formalism.

 The independency assumption takes
the place of the assumption of non-interacting particles, which is
a standard one in statistical mechanics.
Very well known consequences of the assumption of non-interacting
particles, as well as of the independency assumption, are that the
total energy of the system is the sum of the energies of the
particles and that the partition function of the system is the
product of the partition functions of the individual particles.
Identical distribution seems to be a physically sound assumption
for systems of identical particles at the thermal equilibrium even
if, in principle, this assumption is an abstract one that could
not limit the system model to systems at the thermal equilibrium.
Note that the independency assumption makes it possible that two
or more particles occupy the same quantum state. Therefore, in the
case of indistinguishable quantum particles, our model applies
only to systems of bosons, for instance, systems of Helium-4
atoms, not to systems of fermions.

It is worth pointing out that, while from the one hand our model
is genuinely quantum because it captures the randomness that is
inherently present in the result of the quantum measurement, from
the other hand the i.i.d. assumption, that reflects Bob's
ignorance about the interactions that the system had in the past,
rules out of the picture the entanglement between the particles of
the system and between the system and the surrounding environment
from which the system is separated before being brought to Bob's
attention. This is not in contradiction with the two papers
\cite{typpopescu,typgold}, where the entanglement between the
system and the surrounding environment is of fundamental
importance in establishing a complete description of quantum
thermodynamics, see also the recent book \cite{deffner}. Actually,
our scope here is narrower, because this entanglement is not our
(Bob's) concern. As such, the mentioned entanglement is in this
paper an aspect of the experiment that is not specified. This
approach is supported by the following quotation from Lavis
\cite{lavis}:

{\em It seems most reasonable to define the probability as that of
the outcome of a particular experiment with the object in
question, in circumstances where some aspect of the test is
incompletely specified.}
\newline In our model, the measurement destroys the unspecified
entanglement between the system of interest and the environment
from which the system is separated, shifting it to the
entanglement between the system of interest and the measurement
apparatus, which again is neither specified nor it is our concern
because the next experimenter will again ignore the past
interactions of the system. A recent investigation of the
measurement state as an entangled state can be found in
\cite{hobson}.

\section{Distribution of the occupancy numbers}

Let ${\cal N}_l$ be the random number of $D$-tuples of $\bar{{\cal
L}}$ equal to $l$:
\begin{equation}{\cal N}_l \stackrel{\text{def}}{=}
\sum_{i=1}^{N} \delta(l -{\cal L}_i),  \ \ \forall \ l \ \in
\{{\cal L}\}, \label{occn}
\end{equation}
\begin{equation} N=\sum_{l \in \{{\cal L}\}}{\cal
N}_l,\label{totaln}
\end{equation}
where the constraint (\ref{totaln}) is understood in what follows
and $\delta(\cdot)$ is the $D$-dimensional indicator function:
\begin{align} \delta(l)=\delta(l_1,l_2,\cdots l_D)
=\left\{
\begin{array}{cc}  1, &   l_1=l_2=\cdots l_D=0, \\
0,  &  \mbox{elsewhere}.
\end{array} \right.
\nonumber \end{align} The above ${\cal N}_l$ is the  {\em random
occupancy number} of the $l$-th quantum state.  Let the number of
quantum states (the number of $D$-tuples) allowed to one particle
be $|{\cal L}|$ and let $\bar{{n}}$ denote the vector that spans
the support set of the $|{\cal L}|$ random occupancy
numbers.\footnote{When the number of quantum states allowed to the
particle is infinite, to obtain a vector of occupancy numbers with
a finite number of elements we truncate the set $\{{\cal L}\}$ to
a subset $\{{\cal L}'\}$ with finite number of elements. The
truncation is such that
\[Pr({\cal L} \not\in \{{\cal L}'\})  \]
is small enough for our purposes.} Hereafter, the statistical
ensemble $\{p_{\bar{\cal N}}(\bar{n}), \bar{\cal N}\}$ is referred
to as the ensemble of the {\em random occupancy macrostates} or,
in short, {\em random macrostates}. As with microstates, also with
macrostates we will omit the adjective random when the context
makes it redundant.

A consequence of the i.i.d. assumption made in the previous
section is that the probability distribution of the random vector
$\bar{\cal N}$ is the multinomial distribution:
\begin{align}
\hspace{-0.2cm}p_{\bar{\cal N}}(\bar{n})= W(\bar{n})\prod_{l \in
\{{\cal L}\}}(p_{\cal L}(l))^{{n}_l}, \label{md}\end{align} where
$W(\bar{{{n}}})$ is the multinomial coefficient
\begin{equation}
W(\bar{{n}})=\frac{N!}{\prod_{l \in \{{\cal
L}\}}{n}_l!}.\label{shmulti}
\end{equation}
As $N \rightarrow \infty$, by the law of large numbers the random
distribution $\{N^{-1}{\cal N}_l\}$ with multinomially distributed
vector $\bar{\cal N}$ converges to $\{p_{\cal L}(l)\}$, the rate
of convergence being today active area of research, see
\cite{agrawal}. In any case, as we claim in the abstract, the only
condition that is necessary for the vector $\bar{\cal N}$ to be
multinomially distributed is that the random quantum numbers are
i.i.d., independently of the number of particles $N$.

To our best knowledge, it is observed here for the first time that
the distribution of the occupancy numbers of the quantum states is
the multinomial distribution given in (\ref{md}). Although the
multinomial coefficient is pervasive in statistical mechanics
since the times of Boltzmann, the author has been not able to find
the term $\prod_{l \in \{{\cal L}\}}(p_{\cal L}(l))^{n_l}$ that
multiplies the multinomial coefficient in (\ref{md}) in the open
literature, at least with the meaning that it has in (\ref{md}).
It should be said that the multinomial distribution is found very
often in the literature, but not with the meaning that is has
here. An exception is eqn. (8) of 6.2 of \cite{pathria}, where the
multinomial distribution appears in a calculation based on the
multinomial theorem, but the authors of \cite{pathria} seem not to
give to the multinomial distribution any other meaning than that
of a term involved in a mere algebraic manipulation. After, in
contrast to our claim, in the comment to eqn. 6.3.12, the authors
of \cite{pathria} claim that the distribution of the occupancy
numbers in the canonical ensemble is the Poisson distribution.
However, the occupancy numbers cannot follow the Poisson
distribution if the total number of particles is deterministic, as
it actually is in the canonical ensemble, hence this claim of
\cite{pathria} is not compatible with the constraint on the total
number of particle of eqn. 6.2.3 of the same book. In
\cite{niven,safra} and in II.5 of \cite{feller} the multinomial
distribution is considered by taking for $p_{\cal L}$ the uniform
distribution, but the uniform distribution cannot be the
distribution of the occupancy numbers of quantum states. The
multinomial distribution is also used in \cite{swe,dellanna} to
describe the occupancy of space, which again is not the occupancy
of quantum states.

The deterministic transformation (\ref{occn}) from $\bar{\cal L}$
to $\bar{\cal N}$ is such that
\begin{align}  \hspace{-0.3cm}p_{\bar{\cal N},\bar{\cal L}}(\bar{n},\bar{l})
=\left\{
\begin{array}{cc}  p_{\bar{\cal L}}(\bar{l})=\prod_{l \in
\{{\cal L}\}}(p_{\cal L}(l))^{{n}_l}, &   \bar{l} \in
\{\bar{l}(\bar{n})\}, \\
0,  &  \bar{l} \not\in \{\bar{l}(\bar{n})\},
\end{array} \right.
\label{ze2} \end{align}
\begin{align}  p_{\bar{\cal N}|\bar{\cal L}}(\bar{n}|\bar{l})
=\left\{
\begin{array}{cc} 1, &   \bar{l} \in
\{\bar{l}(\bar{n})\}, \\
0,  &  \bar{l} \not\in \{\bar{l}(\bar{n})\},
\end{array} \right.
\label{ze3} \end{align}
\begin{align}
\{\bar{l}(\bar{n})\}=\{\bar{l}_1(\bar{n}),
 \bar{l}_2(\bar{n}),
 \cdots, \bar{l}_{W(\bar{n})}(\bar{n})\},
 \label{permset}
\end{align} where the familiar notation is used for the conditional
probability and for the joint probability, $\bar{l}_1(\bar{n})$ is
a vector whose occupancy numbers are the elements of $\bar{n}$,
the subscript $_i$ indicates the $i$-th distinct permutation of
the $N$ $D$-tuples of $\bar{l}_1(\bar{n})$,\footnote{The
permutation of two particles with the same quantum $D$-tuple is
not a distinct permutation.} and the number of elements
$W(\bar{n})$ of the set of distinct permutations is the
multinomial coefficient (\ref{shmulti}). From (\ref{md}) and
(\ref{ze2}) one promptly recognizes that microstates are {\em
conditionally equiprobable} given the macrostate:
\begin{align}  p_{\bar{\cal L}|\bar{\cal N}}(\bar{l}|\bar{n})
=\left\{
\begin{array}{cc} (W(\bar{n}))^{-1}, &   \bar{l} \in
\{\bar{l}(\bar{n})\}, \\
0,  &  \bar{l} \not\in \{\bar{l}(\bar{n})\}.
\end{array} \right.
\label{condequi} \end{align} At the light of conditional
equiprobability, we see that the i.i.d. assumption conveniently
synthesizes the two standard assumptions of non-interacting
particles and of equally probable microstates in the
microcanonical ensemble. The assumption of non-interacting
particles is mapped onto the factored form of the probability
distribution $p_{\bar{\cal L}}(\bar{l})=\prod_{i=1}^{N}p_{{\cal
L}}(l_i)$, while equiprobability of microstates in the
microcanonical ensemble is demonstrated by deriving
(\ref{condequi}) from the i.i.d. assumption.

Conditional equiprobability is interpreted by saying that, after
that the measurement has detected the occupancy macrostate,
microstates belonging to the known occupancy macrostate are
equiprobable. In our approach, the microcanonical ensemble is
therefore the statistical ensemble $\{p_{\bar{\cal L}|\bar{\cal
N}}(\bar{l}|\bar{n}),{\bar{\cal L}}\}$ and a system of the
microcanonical ensemble is a system whose occupancy numbers
$\bar{n}$ are known to the experimenter through the measurement.
Then, the canonical ensemble $\{p_{\bar{\cal
L}}(\bar{l}),{\bar{\cal L}}\}$ is the weighted union of all the
microcanonical ensembles, the weights being the probabilities
$\{p_{\bar{\cal N}}(\bar{n})\}$. This interpretation, together
with the already discussed lack of entanglement between the
environment and our canonical ensemble, narrows the scope of our
microcanonical ensemble, making it, in this narrow sense,
compatible with the objections that are raised against the
microcanonical ensemble approach in \cite{typpopescu},
\cite{deffner}.

Note that our approach is self-consistent whichever is the number
of particles. In the case of one particle, the vector of the
occupancy numbers is populated by only one non-zero entry, and,
after the measurement, the quantum state of the system is fully
known. In the thermodynamic limit, the {\em relative} randomness
of the occupancy numbers, when represented by the standard
deviations of the random entries of the vector $N^{-1} \bar{\cal
N}$, becomes vanishingly small with $N^{-1/2}$, but at the same
time the {\em absolute} randomness of the random entries of
$\bar{\cal N}$ grows with $N^{1/2}$. Therefore the relative
knowledge about the system gained from measurement of the
occupancy numbers becomes smaller and smaller as the size of the
system increases, but, at the same time, the absolute knowledge
increases with the size of the system, despite the vanishingly
small relative randomness of the occupancy numbers. In the next
section, we will show that this knowledge is quantified by the
information-theoretic mutual information, which turns out to be
equal to the Shannon entropy of the multinomial distribution.

\section{Entropy and mutual information}

The random entropy of ${\cal X}$, or {\em surprise}, denoted
$H({\cal X})$, is a random variable defined by the following
deterministic function of the random ${\cal X}$:
\begin{equation}H({\cal X})\stackrel{\text{def}}{=}-k\log(p_{\cal X}({\cal
X})) \geq 0,\label{entropydef} \end{equation} where $\log(\cdot)$
is the natural logarithm, $k>0$ is a constant that depends on the
context, and $H({\cal X})=0$ only when the distribution $p_{\cal
X}$ is an indicator function, hence when ${\cal X}$ is non-random.
Note that only values of the random variable that occur with zero
probability can cause infinite values of minus the logarithm.
These values can be excluded to all the practical purposes from
the support set of the random variable. The random conditional
entropy of ${\cal X}$ given ${\cal Y}$ is
\begin{equation}H({\cal X}|{\cal Y})\stackrel{\text{def}}{=}
-k\log(p_{{\cal X}|{\cal Y}}({\cal X}|{\cal Y}))
,\label{condentropydef}
\end{equation} where the joint random variable $({\cal X},{\cal
Y})$ is drawn from the joint ensemble with joint probability
distribution $p_{{\cal X},{\cal Y}}$. Here, $H({\cal X}|{\cal
Y})=0$ only when ${\cal X}$ is known given ${\cal Y}$, that is,
when ${\cal X}$ is a deterministic transformation of ${\cal Y}$.
In physics, $k$ is Boltzmann's constant $k_B$, $k_B=1.31 \cdot
10^{-23}$ J/K, while, in information and communication theory,
\[k=\frac{1}{\log(2)},\]
or, equivalently, one takes the base-2 logarithm and $k=1$. In
what follows,  we drop the multiplicative constant, expressing
random entropy and random conditional entropy in $k_B$ units. Note
that we completely skip the notion of phase space in the
definition of (random) entropy and, with it, the need of
low-temperature approximations that are inherent in the standard
phase space approach to physical entropy, see e.g.
\cite{psapprox}.

For the multinomial random variable we have
\begin{align}H(\bar{\cal N})&=-\sum_{l \in \{{\cal L}\}}{\cal N}_l \log(p_{{\cal L}
}(l))-\log\left(W(\bar{\cal N})\right) \label{3dexactr} \\
&=-\sum_{i=1}^N \log(p_{{\cal L} }({\cal
L}_i))-H(\bar{\cal L}|\bar{\cal N}) \label{condint1} \\
&=H(\bar{\cal L})-H(\bar{\cal L}|\bar{\cal N})
\label{randomi},\end{align} where in (\ref{condint1}) we
substitute (\ref{condequi}) and in (\ref{randomi}) we use the
i.i.d. assumption (\ref{indep}). Equation (\ref{3dexactr}) gives
an explicit and exact formula based on the distribution $p_{\cal
L}$ for the physical entropy, in the sense of \cite{plesch}, of
the occupancy macrostate of the system randomly drawn from the
ensemble.

For $N=1$, $\bar{\cal L}$ consists of the only entry ${\cal L}_1$
which is known given $\bar{\cal N}$, therefore $H(\bar{\cal
L}|\bar{\cal N})=0$ and we have
\[H(\bar{\cal N})=H(\bar{\cal L}),\]
while for $N \rightarrow \infty$ we have the limit
\begin{align} \lim_{N
\rightarrow \infty} \frac{H({\bar{\cal L}})-H({\bar{\cal
L}|\bar{\cal N}})}{N}
 &=\lim_{N
\rightarrow \infty} \frac{ H({\bar{\cal N}})}{N}
 =0 \ \ \mbox{in probability,}\nonumber
\end{align}
where {\em in probability} means that the limit is zero for the
all the vectors $\bar{\cal L}$ that occur with non-zero
probability. As expected from the discussion that concluded the
previous section, the above limit shows that, for $N \rightarrow
\infty$,  the randomness of $\bar{\cal N}$ does not impact random
entropy per particle and random conditional entropy per particle
of the microstate, and, at the same time, it shows that the random
entropy per particle of the macrostate becomes vanishingly small
in all the systems of the ensemble that occur with non-zero
probability. This happens because random entropy per particle and
random conditional entropy per particle are impacted by the {\em
relative} randomness of $\bar{\cal N}$, which becomes vanishingly
small in the thermodynamic limit. We remark again that, despite
random entropy per particle and random conditional entropy per
particle of the microstate tend to become equal between them, when
the random entropy of the macrostate is the concern we must
consider the absolute number of particles, hence the {\em
absolute} randomness of $\bar{\cal N}$, that is reflected in the
absolute difference $H({\bar{\cal L}})-H({\bar{\cal L}|\bar{\cal
N}})$ in (\ref{randomi}), which of course grows with $N$.

To characterize the entropy of the canonical ensemble, that we
identify with the variable of state commonly called {\em entropy}
in thermal physics, we consider the expectations over the
ensemble. Specifically, the Shannon-Gibbs entropy $H_{\cal X}$ and
the conditional Shannon entropy $H_{{\cal X}|{\cal Y}}$ are the
expectations of the corresponding random entropies:
\[H_{\cal X} \stackrel{\text{def}}{=}
\braket{H({\cal X})}\stackrel{\text{def}}{=} -\sum_{x \in \{{\cal
X}\}}p_{\cal X}(x)\log(p_{\cal X}(x)),\]
\begin{align}H_{{\cal X}|{\cal
Y}}&\stackrel{\text{def}}{=}\braket{H({{\cal X}|{\cal
Y}})}\nonumber
\\ & \stackrel{\text{def}}{=}-\sum_{x \in \{{\cal X}\}}\sum_{y \in
\{{\cal Y}\}}p_{{\cal X},{\cal Y}}(x,y)\log(p_{{\cal X}|{\cal
Y}}({x}|y)), \nonumber
\end{align}
where we use the angle brackets to denote the expectation with
respect to the random variables (the calligraphic characters) that
are in the argument of the deterministic function inside the angle
brackets and we put $0\log(0)=0$. A straightforward consequence of
the i.i.d. assumption is that the Shannon-Gibbs entropy of the
random microstate is
\[H_{\bar{\cal L}}=NH_{{\cal L}},\]
therefore the Shannon entropy of the multinomial distribution,
that is, of the occupancy macrostate, is
\begin{align}
H_{\bar{\cal N}}&=H_{\bar{\cal L}}- H_{\bar{\cal L}|\bar{\cal N}}
\label{exbol} \\ &=H_{\bar{\cal L}}-\braket{\log(W(\bar{\cal N}))}
\nonumber
\\ &=NH_{\cal L}-\log(N!)+\sum_{l \in \{{\cal
L}\}}\braket{\log({\cal N}_{l}!)},\label{3dexact}
\end{align}
where
\begin{align}& \braket{
\log({\cal N}_l!)}
=\sum_{n=0}^{N}\left(
\begin{array}{c} N \\
n\\
\end{array} \right)
(p_{\cal L}(l))^n(1-p_{\cal L}(l))^{N-n} \log(n!),
\nonumber\end{align} see \cite{me}, see \cite{mahdi} for the
calculation of the expectation in integral form. The sum in
(\ref{3dexact}) seems to be overlooked in the entire literature of
statistical mechanics. This sum, coming from the denominator of
the multinomial coefficient, accounts for the probability that two
or more particles are found in the same quantum state, hence its
nature is clearly quantistic.

The entropic equality (\ref{exbol}) can be found in \cite{zupa},
where the authors call {\em the mean value of the Boltzmann
entropy} our $H_{\bar{\cal L}|\bar{\cal N}}$ and call {\em the
probability distribution of the macroscopic state of the system}
our $p_{\bar{\cal N}}(\bar{n})$. However, the authors of
\cite{zupa} seem not to be aware that their mean value of the
Boltzmann entropy is a conditional entropy. As a matter of fact,
they do not make any explicit use of conditional probability
distributions. Also, they claim but don't prove that the
probability distribution of the occupancy numbers is multinomial,
neither they use the multinomial distribution, in fact they
consider the Gaussian approximation in place of the exact
multinomial distribution. Another big difference with our approach
is that the authors of \cite{zupa} {\em assume} that microstates
are conditionally equiprobable given the macrostate, while we {\em
prove} conditional equiprobability from the i.i.d. assumption.

The mutual information between ${\cal X}$ and ${\cal Y}$, denoted
$I_{{\cal X};{\cal Y}}$, is defined as \begin{equation}I_{{\cal
X};{\cal Y}}\stackrel{\text{def}}{=}H_{{\cal X}}-H_{{\cal X}|{\cal
Y}}= H_{{\cal Y}}-H_{{\cal Y}|{\cal X}} \geq 0, \label{infodef}
\end{equation}
where the second equality, which motivates the adjective {\em
mutual} in front of {\em information}, follows from Bayes rule,
and $I_{{\cal X};{\cal Y}}=0$ only when ${\cal X}$ and ${\cal Y}$
are independent, see chapter 2 of \cite{cover}. From (\ref{exbol})
we have
\begin{equation} I_{\bar{\cal L};\bar{\cal
N}}=H_{\bar{\cal L}}-H_{\bar{\cal L}|\bar{\cal N}}=H_{\bar{\cal
N}}. \label{info}
\end{equation}
Note that, while in the general case $({\cal X},{\cal Y})$ can be
such that
\[p_{{\cal X}}({\cal X})>p_{{\cal X}|{\cal
Y}}({\cal X}|{\cal Y}),\] hence it can happen with non-zero
probability that
\[H({\cal X})-H({{\cal X}|{\cal
Y}}) < 0,\] in our case we always have
\[H(\bar{\cal L})-H(\bar{\cal L}|\bar{\cal N})=H(\bar{\cal N})-H(\bar{\cal N}|\bar{\cal L})
=H(\bar{\cal N})\geq 0\] where the second equality holds because,
since $\bar{\cal N}$ is a deterministic function of $\bar{\cal
L}$, $H(\bar{\cal N}|\bar{\cal L})=0$ whichever is $\bar{\cal L}$.
We see therefore that the measurement of the occupancy numbers
gives non-negative contribution to the mutual information
(\ref{info}) for all the systems of the ensemble that can come out
with non-zero probability, while inequality (\ref{infodef})
guarantees non-negative information between two generic random
variables only after that the expectation over the ensemble has
been taken.



 A comment is in order
about the relationship between information and physical
distinguishability/indistinguishability of particles of the same
species. We adhere to the following widely accepted definition of
distinguishability/indistinguishability. Classical particles are
said to be distinguishable when their motion can be tracked, while
quantum particles are said to be distinguishable when their wave
functions do not overlap. This can be the case of the $N$ vertexes
of a lattice in a solid, or of $N$ distinct containers each one
containing one of the $N$ particles that together constitute the
$N$-particle system. On the opposite, when it is impossible to
track particles' motion or, in the quantum case, when the wave
functions of particles overlap, they are said to be {\em
indistinguishable}. This can be the case when particles share the
same region of space, for instance, because they are in the same
container. Indistinguishability is independent of particles'
quantum state, therefore two atoms of Helium-4 in the same
container are for us indistinguishable bosons, even if the
measurement collapses them in two different quantum states. The
mutual information (\ref{info}) is a measure of how much
information can be encoded by an encoder, call it Alice, when she
prepares an occupancy macrostate randomly drawn from the
statistical ensemble of macrostates, and detected without errors
by a decoder, call it Bob, when he analyzes the preparation with
the aim of detecting which specific member of the ensemble has
been prepared.\footnote{The characters Alice and Bob and the term
{\em preparation} are intentionally drawn from the language of
quantum information theory and have the same meaning that they
have in that context, see e.g. \cite{ikemike}.} Bob can make
errorless detection of the macrostate by projecting the system
onto its energy eigenbasis and then by applying the deterministic
transformation (\ref{occn}) to the result $\bar{\cal L}$ of the
projective measurement. This procedure is feasible both with
distinguishable and with indistinguishable particles, because the
result of the deterministic transformation (\ref{occn}) is
independent of where particles are found, that is, it is
independent of particles' indexes. In the case of distinguishable
particles, another communication procedure is feasible. In this
procedure, Alice and Bob agree about an unambiguous indexing of
the $N$ particles. This can be done, for instance, when particles
occupy distinct regions of space. Now Alice can make $N$
independent mappings, one for each particle, each of which carries
$H({\cal L})$ units of information, and Bob can successfully
detect information particle-by-particle, extracting from the
system $H(\bar{\cal L})$ units of information. However, when
particles are indistinguishable, any agreement between Alice and
Bob about indexing fails. Due to lack of indexing, Bob has not
access to the microstate of the prepared macrostate. Since the
$W(\bar{\cal N})$ microstates whose union form the macrostate
$\bar{\cal N}$ are equiprobable, compared to the case where
particles are indexed, Bob has not access to $\log(W(\bar{\cal
N}))$ units of information, whose expectation over the canonical
ensemble is just $\braket{\log(W(\bar{\cal N}))}=H_{\bar{\cal
L}|\bar{\cal N}}$ units of information.


\section{Quantum-mechanical representation}

Although the absence of entanglement makes it possible to treat
the quantum system by classical random variables, the quantum
mechanical formalism that we are going to introduce will put light
on the two quantum models behind $H_{\bar{\cal N}}$ and
$H_{\bar{\cal L}}-H_{\bar{\cal L}|\bar{\cal N}}$ and on the
equivalence between them.

 Let the set of energy
eigenstates $\{\ket{{\bar{l}}}\}$ form a complete orthonormal
basis of the Hilbert space of the system, with \begin{equation}
\ket{\bar{l}}= \ket{l_1,l_2, \cdots l_N}=\ket{l_1}_1\otimes
\ket{l_2}_2\otimes \cdots \ket{l_N}_N,\label{tp}\end{equation}
where $\otimes$ denotes the tensor product and
\[\{\ket{{{l}}}_i\},
\ \ {{l}} \in \{{{\cal L}}\}, \ \ i=1,2,\cdots,N,\] form an
orthonormal basis for the $D$-dimensional Hilbert subspace of the
$i$-th particle:
\[\ket{l}_i=
\ket{l_1,l_2, \cdots l_D}_i= \ket{l_{1}}_{i,1}\otimes
\ket{l_{2}}_{i,2}\otimes \cdots \ket{l_{D}}_{i,D}.\]

Two quantum mechanical representations of the occupancy macrostate
are hereafter considered. The first one is based on the mixed
state represented by the statistical ensemble of density operators
$\{p_{\bar{\cal N}}(\bar{n}), \hat{\rho}(\bar{\cal N})\}$ with
\begin{align}
\hat{\rho}(\bar{n})=\sum_{\bar{l} \in \{\bar{\cal
L}\}}p_{\bar{\cal L}|\bar{\cal N}}(\bar{l}|\bar{n})
\ket{{\bar{l}}}\bra{{\bar{l}}}, \ \ \forall \ \bar{n} \in
\{\bar{\cal N}\}.
 \label{permset}
\end{align} Substituting
(\ref{condequi}) in (\ref{permset}), we write
\begin{align}
\hat{\rho}(\bar{n})=\frac{1}{W(\bar{n})}\sum_{i=1}^{W(\bar{n})}
\ket{\bar{l}_i(\bar{n})}\bra{\bar{l}_i(\bar{n})}, \ \ \forall \
\bar{n} \in \{\bar{\cal N}\}.
 \label{permset2}
\end{align}
The expectation over the set of classical random vectors
$\{\bar{\cal N}\}$ gives
\begin{align}\hat{\rho}&=\braket{\hat{\rho}(\bar{\cal N})} \nonumber \\
&=\sum_{\bar{n} \in \{\bar{\cal N}\}} p_{\bar{\cal{N}}}(\bar{n})
\hat{\rho}(\bar{n}) \nonumber
\\ &=\sum_{\bar{n} \in \{\bar{\cal
N}\}}p_{\bar{\cal{N}}}(\bar{n})\sum_{\bar{l} \in \{\bar{\cal
L}\}}p_{\bar{\cal L}|\bar{\cal N}}(\bar{l}|\bar{n})
\ket{{\bar{l}}}\bra{{\bar{l}}} \nonumber \\ &=\sum_{\bar{l} \in
\{\bar{\cal L}\}}p_{\bar{\cal L}}(\bar{l})
\ket{{\bar{l}}}\bra{{\bar{l}}}.\label{mixedsystem2} \end{align}
Using (\ref{tp}) and the i.i.d. assumption (\ref{indep}), the
density operator $\hat{\rho}$ can be factored into the density
operators of the $N$ particles:
\begin{align}\hat{\rho}&=  \hat{\rho}_1 \otimes \hat{\rho}_2 \otimes \cdots \otimes
\hat{\rho}_N,\label{seppart}
\end{align}
where
\[\hat{\rho}_i=\sum_{l \in \{{\cal L}\}} p_{{\cal
L}}({l})\ket{{{l}}}_{i}\bra{{{l}}}_i, \ \ i=1,2, \cdots,N,\] is
the density operator of the i-th particle.

The second quantum mechanical representation is based on the
statistical ensemble of pure states $\{p_{\bar{\cal N}}(\bar{n}),
\ket{\phi(\bar{\cal N})}\bra{\phi(\bar{\cal N})}\}$ with
\begin{align}\ket{\phi(\bar{n})}&=\sum_{\bar{l}\in \{\bar{\cal L}\}}
\sqrt{p_{\bar{\cal L}|\bar{\cal
N}}(\bar{l}|\bar{n})}\ket{\bar{l}}\nonumber \\ &=
\sqrt{\frac{1}{W(\bar{n})}} \sum_{i=1}^{W(\bar{n})}
\ket{\bar{l}_i(\bar{n})}, \ \ \forall \ \ \bar{n} \in \{\bar{\cal
N}\},\label{phin}\end{align} which is the standard textbook
representation of the pure state of a system of $N$ bosons, see
e.g. \cite{kardar}. One more time we remark that, compared to the
standard approach, where equiprobability is postulated, in our
approach equiprobability is a consequence of conditional
equiprobability that, through the multinomial distribution,
descends from the i.i.d. assumption. The expectation over
$\{\bar{\cal N}\}$ provides the density operator
\begin{equation}\hat{\rho}'=\braket{\ket{\phi(\bar{\cal N})}\bra{\phi(\bar{\cal N})}}=\sum_{\bar{n} \in \{\bar{\cal N}\}}
p_{\bar{\cal{N}}}(\bar{n})
\ket{\phi(\bar{n})}\bra{\phi(\bar{n})},\label{rhoprime}
\end{equation}
which is not (\ref{mixedsystem2}), neither has the same
eigenvalues. Nonetheless, we hereafter show that $\hat{\rho}$ and
$\hat{\rho}'$ are equivalent to our purposes.


Thanks to the  orthogonality of the pure states that populate the
density operators $\hat{\rho}$, $\hat{\rho}'$ and
$\{\hat{\rho}({\bar{n}})\}$, the Von Neuman entropy $S(\cdot)$,
\[S(\hat{\sigma})\stackrel{\text{def}}{=}-\mbox{Tr}\left(\hat{\sigma}
\log(\hat{\sigma})\right),\] is equal to the Shannon entropy of
the classical random variable that characterizes the density
operator, therefore
\[S(\hat{\rho})=H_{\bar{\cal L}},\]
\begin{equation}S(\hat{\rho}(\bar{n}))=-\sum_{\bar{l} \in \{\bar{\cal
L}\}}p_{\bar{\cal L}|\bar{\cal
N}}(\bar{l}|\bar{n})\log(p_{\bar{\cal L}|\bar{\cal
N}}(\bar{l}|\bar{n})),\label{sub} \end{equation}
\[S(\hat{\rho}')=H_{\bar{\cal N}}.\]
Let us now consider the case where the statistical ensemble of
preparations is $\{p_{\bar{\cal N}}(\bar{n}),\hat{\rho}(\bar{\cal
N})\}$. The Holevo upper bound $\chi$ above the accessible
information, see chapter 12 of \cite{ikemike} for the definition,
is
\begin{align}\chi &\stackrel{\text{def}}{=}
S(\hat{\rho})-\sum_{\bar{n} \in \{\bar{\cal N}\}} p_{\bar{\cal
N}}(\bar{n})S(\hat{\rho}({\bar{n}})) \nonumber
\\& =H_{\bar{\cal L}}- H_{\bar{\cal L}|\bar{\cal N}} =
I_{\bar{\cal L};\bar{\cal N}}, \label{joint}
\end{align}
where the right hand side of the first equality of (\ref{joint})
is obtained by substituting (\ref{sub}) and by using $p_{\bar{\cal
L}|\bar{\cal N}}(\bar{l}|\bar{n})p_{\bar{\cal
N}}(\bar{n})=p_{\bar{\cal L},\bar{\cal N}}(\bar{l},\bar{n})$. If
the statistical ensemble of preparations is $\{p_{\bar{\cal
N}}(\bar{n}), \ket{\phi(\bar{\cal N})}\bra{\phi(\bar{\cal N})}\}$
we have
\begin{align}\chi' &\stackrel{\text{def}}{=}
S(\hat{\rho}')-\sum_{\bar{n} \in \{\bar{\cal N}\}} p_{\bar{\cal
N}}(\bar{n})S(\ket{\phi(\bar{n})}\bra{\phi({\bar{n}})}) \nonumber
\\& =S(\hat{\rho}') \label{zp} \\
&=H_{\bar{\cal N}}=I_{\bar{\cal L};\bar{\cal N}},  \nonumber
\end{align}
where (\ref{zp}) holds because the Von Neumann entropy of any pure
state is zero. This shows that the two statistical ensembles are
equivalent to what concerns the information that can be
encoded/decoded into/from the preparation. Actually, whichever is
the density operator of the preparation among
$\hat{\rho}(\bar{n})$ and
$\ket{\phi(\bar{n})}\bra{\phi(\bar{n})}$, nothing changes for Bob.
He operates the measurement by the complete set of projectors
$\{\ket{{\bar{l}}}\bra{{\bar{l}}}\}$, with arbitrary indexing of
particles' subspaces. The measurement leaves the system in the
eigenstate $\ket{\bar{l}}$ with probability $(W(\bar {n}))^{-1}$:
\begin{align}\braket{\bar{l}|\hat{\rho}(\bar{n})|\bar{l}}&=
\sum_{\bar{l}' \in \{\bar{\cal
L}\}}p_{\bar{\cal{L}}|{\bar{\cal{N}}}}({\bar{l}'|\bar{n}})\braket{\bar{l}|\bar{l}'}\braket{\bar{l}'
|\bar{l}}\nonumber
\\ &=p_{\bar{\cal{L}}|{\bar{\cal{N}}}}({\bar{l}|\bar{n}}) \nonumber \\ &=\left\{
\begin{array}{cc} (W(\bar{n}))^{-1}, &   \bar{l} \in
\{\bar{l}(\bar{n})\}, \\
0,  &  \bar{l} \not\in \{\bar{l}(\bar{n})\},
\end{array} \right.\nonumber
\end{align}
\begin{align}
|\braket{\bar{l}|\phi(\bar{n})}|^2&= \left|\sum_{\bar{l}' \in
\{\bar{\cal
L}\}}\sqrt{p_{\bar{\cal{L}}|{\bar{\cal{N}}}}({\bar{l}'|\bar{n}})}
\braket{\bar{l}|\bar{l}'}\right|^2\nonumber
\\ & =p_{\bar{\cal{L}}|{\bar{\cal{N}}}}({\bar{l}|\bar{n}}) \nonumber \\ &=\left\{
\begin{array}{cc} (W(\bar{n}))^{-1}, &   \bar{l} \in
\{\bar{l}(\bar{n})\}, \\
0,  &  \bar{l} \not\in \{\bar{l}(\bar{n})\}.
\end{array} \right.\nonumber
\end{align}
Then he performs errorless detection of the macrostate from
particles' states that come out from the measurement by the
index-independent procedure already discussed, which is the same
in the two cases.

\section{Entropy of the ideal gas at the thermal equilibrium}

For systems at the thermal equilibrium, the distribution $p_{{\cal
L}}$  is the one-particle Boltzmann distribution:
\begin{equation}p_{{\cal
L}}({l})=Z^{-1}e^{-\beta \eta({l})},\label{boltz0}
\end{equation}
where  $Z$ is the one-particle partition function,
\begin{equation}Z=\sum_{l \in \{{\cal L}\}}e^{- \beta \eta(l)},
\label{pf}
\end{equation} $\eta(l)$ is
the energy eigenvalue associated to the state number $l$ and
$\beta=(k_B T)^{-1}$, where $T$ is the temperature in Kelvin
degrees of the heat bath that thermalizes the system.
 By the i.i.d.
assumption, also the distribution $p_{\bar{\cal L}}$ is Boltzmann:
\footnote{The Boltzmann distribution maximizes $H_{\bar{\cal L}}$
under the temperature constraint. Due to the sum in
(\ref{3dexact}) the Boltzmann distribution could not maximize
$H_{\bar{\cal N}}$. However, maximization of $H_{\bar{\cal N}}$ is
out of the scope of this paper.  }
\begin{equation}p_{\bar{\cal L}}(\bar{l})=\prod_{i=1}^{N}
p_{{\cal L}}({l}_i)=Z^{-N}e^{-\beta
\sum_{i=1}^{N}\eta(l_i)}.\label{boltz1}
\end{equation}
Using
\begin{equation}\sum_{i=1}^{N}\eta(l_i)=
\sum_{l\in \{{\cal L}\}} n_l\eta(l),\label{twosides}
\end{equation}
in the Boltzmann distribution and the Boltzmann distribution in
the multinomial distribution we get
\[p_{\bar{\cal N}}(\bar{n})=W(\bar{n})Z^{-N}\prod_{l\in \{{\cal L}\}}
e^{-\beta n_l\eta(l)},\]

In the following we consider an ideal monoatomic dilute gas in a
cubic container of side $L$. One particle of the gas is modelled
as a quantum ''particle in a box'' with three degrees of freedom,
whose energy eigenvalues with aperiodic boundary conditions are
\begin{equation}\eta(l)=
(l_x^2+l^2_y+l^2_z) \frac{h^2}{8 m L^2},\ \ l_{x,y,z}
=1,2,\cdots,|{\cal L}_{x,y,z}|,\label{pib}
\end{equation}
where the $D$-tuple $l$ consists of the three quantum numbers
$(l_x,l_y,l_z)$, $l_{x,y,z}$ indicates anyone among
$\{l_x,l_y,l_z\}$, $m$ is the mass of the particle, $h=6.626 \cdot
10^{-34} \ \ \mbox{J} \cdot \mbox{s}$ is the Planck constant, and
we take $|{\cal L}_{x,y,z}|$ large enough to make the probability
of ${\cal L}_{x,y,z}>|{\cal L}_{x,y,z}|$ negligible to our
purposes. For the distribution of the triple of quantum numbers,
we use the energy quantization rule (\ref{pib}) in the Boltzmann
distribution (\ref{boltz0}).

The exact information/entropy formula (\ref{3dexact}) includes two
quantum effects, one impacting the entropy $H_{{\bar{\cal L}}}$
the other impacting the conditional entropy $H_{{\bar{\cal
L}|\bar{\cal N}}}$. The quantum effect that impacts $H_{\bar{\cal
L}}$ is the discrete nature of the sum (\ref{pf}) that defines
the partition function of the Boltzmann distribution.
Approximating the sum (\ref{pf}) over $ \ l=(l_x,l_y,l_z)$ to an
integral, see eqn. 19.54 of \cite{sek}, we get\[Z \approx
\left(\int_0^{\infty}e^{-
 \beta l^2  h^2/8 m L^2}dl\right)^3=\left(\sqrt{\frac{2 \pi m L^2}{
 \beta h^2}}\right)^3,\]
\[\braket{\eta({\cal L})}\approx \frac{3}{2 \beta},\]
leading to the approximation
\begin{align}
H_{{\cal L}}&\approx \frac{3}{2} \left(1+\log \left(\frac{2 \pi m
k_B T L^2}{h^2}\right)\right)  \label{3dapprox1}.
\end{align} The quantum
effect that impacts the conditional entropy $H_{{\bar{\cal
L}|\bar{\cal N}}}$, is represented by the triple sum over
$l=(l_x,l_y,l_z)$ appearing in (\ref{3dexact}). If we neglect this
term, the conditional entropy becomes
\begin{align}
H_{\bar{\cal L}|\bar{\cal N}} & \approx \log(N!), \label{twoterms}
\end{align}
which is the term introduced by Gibbs to make the non-quantized
phase-space (differential) entropy of systems of indistinguishable
particles compatible with his famous paradox. This term, after
more than one hundred years from its appearance, also today is
still object of research and debate, see \cite{sasa,tasaki}.
Using in (\ref{info}) the two approximation above and Stirling's
approximation $\log(N!)\approx N\log(N)-N$ for (\ref{twoterms}) we
obtain the textbook Sackur-Tetrode entropy formula:
\begin{align}
H_{\bar{\cal N}}&\approx N\left(\log \left(
\frac{L^3}{N}\left(\frac{2 \pi m k_B T}{h^2
}\right)^{\frac{3}{2}}\right)+\frac{5}{2} \right).
\label{3dapprox3}
\end{align}

\begin{figure}[!h]
\vspace*{.2cm}
    \centering
    \includegraphics[width=.50\textwidth]{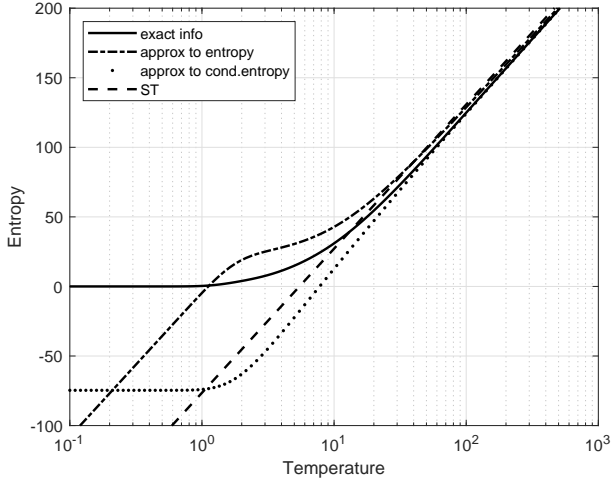}
    \caption{Entropy and its approximations
     for 30 particles with
    three degrees of freedom, quantization rule (\ref{pib}), distribution (\ref{boltz0}).
    Solid line: exact entropy-information.
    Dash-dotted line: approximation obtained using (\ref{3dapprox1}) for $H_{\cal L}$.
        Dotted line: approximation obtained using (\ref{twoterms})
        for $H_{\bar{\cal L}|\bar{\cal N}}$.
        Dashed line: Sackur-Tetrode entropy formula (\ref{3dapprox3}).}
    \label{fig:piab30}
    \vspace*{.3cm}
\end{figure}

\begin{figure}[!h]
\vspace*{.2cm}
    \centering
    \includegraphics[width=.50\textwidth]{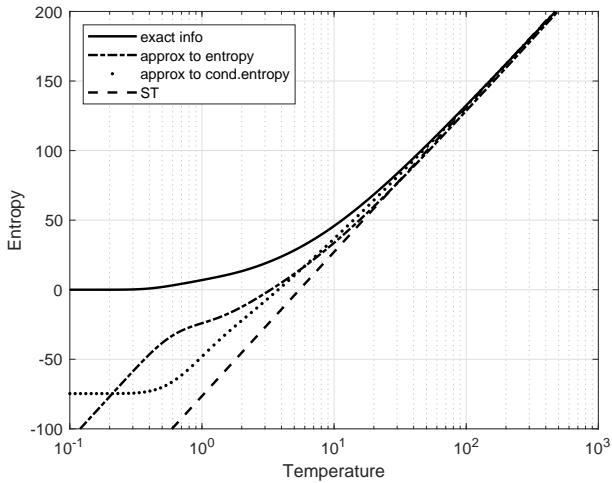}
    \caption{
    Same graphs of Fig.\ref{fig:piab30}, but here we allow for
    $l_{x,y,z}=0$ in (\ref{pib}) and (\ref{pf}).}
    \label{fig:corr30}
    \vspace*{.3cm}
\end{figure}

The exact entropy (\ref{3dexact}) cannot be calculated in closed
form, but its numerical evaluation is feasible. The results of
Figs. \ref{fig:piab30}, \ref{fig:corr30}, report the entropy and
its approximations in $k_B$ units versus temperature in Kelvin
degrees. All the results are obtained for $N=30$ particles, side
of the cubic box $L=10^{-9}$ meters and mass of one particle
$m=1.67 \cdot 10^{-27}$ kilograms.  In our numerical evaluation,
the number $|{\cal L}_{x,y,z}|$ of quantum states on one dimension
considered in the triple sum of equation (\ref{3dexact}) is 10,
which generates a set $\{{\cal L}\}$ of $10^3$ elements, but we
have verified that also $|{\cal L}_{x,y,z}|=5$ does virtually not
impact graphs. For the numerical evaluation of the partition
function we truncate the sum to $|{\cal L}_{x,y,z}|=500$. Also in
this case, we have verified that it is more than enough not to
impact the graphs. We hasten to point out that our aim here is
only to investigate a parametrization that stresses certain
behaviors of the exact information/entropy and of the
approximations already discussed, not to claim that the i.i.d.
assumption and the consequent results are physically sound when
this parametrization is applied to the underlying physical system.

Fig. \ref{fig:piab30} shows that all the approximations discussed
so far, including the Sackur-Tetrode formula, fall below zero at
low temperature, hence they cannot be exact entropy formulas. At
$T=1$ Kelvin, the exact formula is close to zero. This means that
virtually all the particles are in the ground state
$l_x=l_y=l_z=1$. At this temperature, the gap between the exact
formula and (\ref{twoterms}), (\ref{3dapprox3}), is about
$\log(N!)=\log(30!)=74.6$. Since at this temperature $H_{\cal L}
\approx 0$, the approximation obtained by using (\ref{twoterms})
consists only of the term $-\log(N!)$. If, as in the exact
formula, we include the third term of (\ref{3dexact}), we see
that, as expected, at temperature sufficiently low it completely
cancels the term $-\log(N!)$. This term therefore prevents
negative values of the entropy. Below $T=1$ Kelvin, the
approximation obtained by using (\ref{3dapprox1}) follows the same
trend of the Sackur-Tetrode formula, the constant gap between the
two being due to the triple sum in (\ref{3dexact}) which is
included in the approximation based on (\ref{3dapprox1}) while it
is not included in (\ref{3dapprox3}).

Note that the quantization rule (\ref{pib}) excludes from the
support set $l_{x,y,z}=0$, the zero energy level. Technically
speaking, the exclusion of the zero energy level is justified by
observing that,
with zero energy, the solution of the Schr\"{o}dinger equation is
the trivial wave function equal to zero everywhere. Normalization
of its squared modulus fails, leading to the exclusion of the zero
energy eigenvalue from the set of legal eigenvalues. The intuition
behind this technical reason is that if the wave function is zero
everywhere then there is zero probability of finding the particle.
At the same time, if we are sure 100$\%$ that we put it inside the
box before cooling it down to very low temperature, then it must
still be somewhere, leading to the conclusion that zero energy
must be excluded. We observe that there is also another
possibility: the particle really {\em is} inside the box, but our
attempts to find it fail. Actually, if it has zero energy then its
interaction with our instrument is likely to happen with zero
probability. This is compatible with a wave function equal to zero
everywhere and with the fact that we put the particle inside the
box before cooling it down. Including the zero energy level, we
find the results reported in Fig. \ref{fig:corr30}. We don't claim
that our reasoning and, as a consequence, the results of Fig.
\ref{fig:corr30}, although logically conceivable, do physically
make sense. At the same time, to our eyes the graphs displayed in
the two pictures look much better with the zero energy level than
without it. Based on the merely logical reasoning and on the nice
look of the resulting graphs, we decided to show Fig.
\ref{fig:corr30} to the reader.

\section{Conclusions}


The paper has substantiated Jaynes' intuition that the entropy of
a physical system of indistinguishable particles is the
information-theoretic entropy of the macrostate of the system
\cite{jgibbs}, specifically, the entropy of the random occupancy
macrostate. The key observation is that the occupancy numbers of
the quantum states of system's particles that come out from the
random experiment operated by the quantum measurement that
projects the system onto its energy eigenbasis are multinomially
distributed. This observation paves the way to the calculus of
entropy and to its identification with the mutual information
between macrostates and microstates.

A physically counter-intuitive consequence of the exact entropy
formula (\ref{3dexact}) is that the entropy of the ideal gas is
non-extensive, in the sense that
\[H_{\bar{\cal N}}(T, NV, N) \neq NH_{\bar{\cal N}}(T,V,1),\]
also for large $N$, the difference between the two becoming
important at low temperature and/or at high density. Actually, the
Sackur-Tetrode formula (\ref{3dapprox3}) is extensive, but
extensivity arrives only after that quantum effects are washed out
by the high-temperature approximations (\ref{3dapprox1}),
(\ref{twoterms}), and by Stirling's approximation. Playing with
(\ref{3dexact}), we have found
\[H_{\bar{\cal N}}(T, NV, N) - NH_{\bar{\cal N}}(T,V,1) \geq 0\]
with all the parametrizations that we have explored. The
inequality is potentially interesting because it shows that, when
volume is constrained, it is more informative/entropic putting
particles together in the same container than constraining them
inside distinct containers. However, we have not been able to
prove it.

We now briefly sketch how our approach applies to the fall of
entropy that occurs in one-particle quantum Szilard engines when
the volume occupied by the particle is partitioned in two equal
volumes by the insertion of the piston. The complete analysis is
left to future work. Let  ${\cal B}$ be a binary random variable
with distribution
\[p_{{\cal B}}(B=1)=p_{{\cal B}}(B=0)=\frac{1}{2},\]
where ${\cal B}=1$ means that, after the insertion of the piston,
the particle is found on the left half of the volume initially
occupied by the particle. The quantum state ${\cal L}$ of the
particle after the insertion of the piston can be conveniently
expressed by the following joint random variable
\[{\cal L}=({\cal B},{\cal L}'),\]
where ${\cal L}'$ is the quantum state of a particle that occupies
half of the initial volume. Assuming independency between ${\cal
B}$ and ${\cal L}'$, the entropy of the final state is
\begin{align}H_{{\cal L}}&=H_{{\cal
B},{\cal L}'}=H_{{\cal B}}+H_{{\cal L}'}=\log(2)+H_{{\cal L}'},
\label{land}\end{align} where, as expected, the celebrated
$\log(2)$ of Landauer comes from the random variable ${\cal B}$,
hence from the uncertainty about which half of the volume the
measurement will localize the particle in. We have numerically
evaluated the partition function of the Boltzmann distribution
with the parametrization of \cite{aydin}, that is mass of the
particle $m=9.11 \cdot 10^{-31}$ kg, temperature $T=300 $ K, and
one-dimensional box of size $L=20 \cdot 10^{-9}$ m. We obtain that
the entropy of the single particle with one degree of freedom
before the insertion of the piston is $1.988$ in $k_B$ units,
while with size of the one-dimensional box equal to $10 \cdot
10^{-9}$ m, that is, after the insertion of the piston, the
entropy $H_{{\cal L}'}$ in $k_B$ units is $1.243$, leading to the
difference
\[1.988-0.693-1.243=0.052,\]
in excellent agreement with the entropy fall shown in Fig. 3 of
\cite{aydin}, where the result is derived by totally different
means than ours.



\begin{thebibliography}{999}


\bibitem{tribus}
Tribus, Myron, and Edward C. McIrvine. "Energy and information."
Scientific American 225.3 (1971): 179-190.

\bibitem{obsafra}
Safranek, Dominik, et al. "A brief introduction to observational
entropy." Foundations of Physics 51 (2021): 1-20.

\bibitem{neritut} Merhav, Neri. "Statistical physics and information
theory." Foundations and Trends® in Communications and Information
Theory 6.1–2 (2010): 1-212.

\bibitem{land2}
Landauer, Rolf. "Information is a physical entity." Physica A:
Statistical Mechanics and its applications 263.1-4 (1999): 63-67.

\bibitem{mt} Maroney, Owen Jack Ernest. "Information and entropy
in quantum theory." arXiv preprint quant-ph/0411172 (2004).

\bibitem{maroneyrecent}
Maroney, Owen JE, and Christopher G. Timpson. "How is there a
physics of information? On characterizing physical evolution as
information processing." Physical perspectives on computation,
computational perspectives on physics (2018): 103-126.


\bibitem{maxent} Jaynes, E. T. "Information theory and
statistical mechanics." {\em Physical review}, 1957, {\em 4},
620--630.

\bibitem{merhav}  Merhav, Neri. "Physics of the Shannon Limits,"
{\em IEEE Trans. on Inform. Theory,} 2010, {\em 9}, 4274--4285.


\bibitem{cover} Cover, T. M.; and Thomas, J. A.
"Elements of information theory, 2$^{nd} $ edition."  Wiley series
in telecommunications and signal processing, 2006.

\bibitem{spalvieri}
Spalvieri, Arnaldo. "The Shannon-McMillan theorem proves
convergence to equiprobability of Boltzmann's microstates."
Entropy 2021, 23, 899.

\bibitem{landauer}
Landauer, Rolf. "Irreversibility and heat generation in the
computing process." IBM journal of research and development 5.3
(1961): 183-191.



\bibitem{landdemonstration}
Bérut, Antoine, et al. "Experimental verification of Landauer's
principle linking information and thermodynamics." Nature 483.7388
(2012): 187-189.

\bibitem{plenio}
Plenio, Martin B., and Vincenzo Vitelli. "The physics of
forgetting: Landauer's erasure principle and information theory."
Contemporary Physics 42.1 (2001): 25-60.

\bibitem{trends}
Von Spakovsky, Michael R., and Jochen Gemmer. "Some trends in
quantum thermodynamics." Entropy 16.6 (2014): 3434-3470.

\bibitem{itthermoreview}
Goold, John, et al. "The role of quantum information in
thermodynamics - a topical review." Journal of Physics A:
Mathematical and Theoretical 49.14 (2016): 143001.


\bibitem{parrondo}
Parrondo, Juan MR, Jordan M. Horowitz, and Takahiro Sagawa.
"Thermodynamics of information." Nature physics 11.2 (2015):
131-139.










%
%
%



%
%




\bibitem{kardar} Kardar, M.
 "Statistical physics of particles." Cambridge University Press,
2007.

\bibitem{plesch}
Plesch, Martin, et al. "Maxwell's Daemon: Information versus
Particle Statistics." Scientific reports 4.1 (2014): 6995.

\bibitem{typpopescu}
Popescu, Sandu, Anthony J. Short, and Andreas Winter.
"Entanglement and the foundations of statistical mechanics."
Nature Physics 2.11 (2006): 754-758.

\bibitem{typgold}
Goldstein, Sheldon, et al. "Canonical typicality." Physical review
letters 96.5 (2006): 050403.

\bibitem{deffner}
Deffner, Sebastian, and Steve Campbell. Quantum Thermodynamics: An
introduction to the thermodynamics of quantum information. Morgan
and Claypool Publishers, 2019.

\bibitem{lavis}
Lavis, D. A. "The concept of probability in statistical
mechanics." Frontiers of fundamental physics 4 (2001): 293-308.

\bibitem{hobson}
Hobson, Art. "Entanglement and the measurement problem." Quantum
Engineering 2022 (2022): 1-12.

\bibitem{agrawal}
Agrawal, Rohit. "Finite-sample concentration of the multinomial in
relative entropy." IEEE Transactions on Information Theory 66.10
(2020): 6297-6302.

\bibitem{pathria} Pathria, R. K.; Beale, P. D. "Statistical Mechanics,
3$^{rd} $ edition." Elsevier, 2011.

\bibitem{safra}
Safranek, Dominik, Anthony Aguirre, and J. M. Deutsch. "Classical
dynamical coarse-grained entropy and comparison with the quantum
version." Physical Review E 102.3 (2020): 032106.

\bibitem{niven}
Niven, Robert K. "Origins of the combinatorial basis of entropy."
Aip conference proceedings. Vol. 954. No. 1. American Institute of
Physics, 2007.

\bibitem{feller}
Feller, William. "Introduction to Probability Theory Third
Edition, Vol. 1." (1968).



\bibitem{swe}
Swendsen, Robert H. "Statistical mechanics of classical systems
with distinguishable particles." Journal of statistical physics
107.5 (2002): 1143-1166.

\bibitem{dellanna}
Dell'Anna, Luca. "Entanglement properties and ground-state
statistics of free bosons." Physical Review A 105.3 (2022):
032412.









\bibitem{psapprox}
Qiu, Tian, et al. "Quantum corrections to the entropy and its
application in the study of quantum Carnot engines." Physical
Review E 101.3 (2020): 032113.



\bibitem{me}
Kaji, Yuichi. "Bounds on the entropy of multinomial distribution."
2015 IEEE International Symposium on Information Theory (ISIT).
IEEE, 2015.

\bibitem{mahdi}
Cheraghchi, Mahdi. "Expressions for the entropy of binomial-type
distributions." 2018 IEEE International Symposium on Information
Theory (ISIT). IEEE, 2018.

\bibitem{zupa}
\u{Z}upanovi\'{c}, Pa\u{s}ko, and Domagoj Kui\'{c}. "Relation
between Boltzmann and Gibbs entropy and example with multinomial
distribution." Journal of Physics Communications 2.4 (2018):
045002.






\bibitem{ikemike}
Nielsen, Michael A., and Isaac L. Chuang. "Quantum Computation and
Quantum Information." (2010).

\bibitem{sek}
Sekerka, R. F. "Thermal Physics: Thermodynamics and Statistical
Mechanics for Scientists and Engineers." Elsevier, 2015.

\bibitem{sasa}
Sasa, Shin-ichi, et al. "Quasi-static Decomposition and the Gibbs
Factorial in Small Thermodynamic Systems." Journal of Statistical
Physics 189.2 (2022): 31.

\bibitem{tasaki}
Tasaki, Hal. "The best answer to the puzzle of Gibbs about $ N!
$!: A note on the paper by Sasa, Hiura, Nakagawa, and Yoshida."
arXiv preprint arXiv:2206.05513 (2022).


\bibitem{jgibbs}
Jaynes, Edwin T. "The Gibbs paradox." Maximum entropy and bayesian
methods. Springer, Dordrecht, 1992. 1-21.



\bibitem{aydin}
Aydin, Alhun, Altug Sisman, and Ronnie Kosloff. "Landauer's
Principle in a quantum Szilard engine without Maxwell's Demon."
Entropy 22.3 (2020): 294.




%





%
%























%
%
%
%
%
%
%
%
%
%
%
%
%
%
%
%
%
%
%
%
%
%
%
%
%
%






%
%
%
%
%
%
%
%
%
%
%
%
%
%
%
%
%
%
%
%
%
%
%
%
%
%
%
%
%
%
%
%
%
%
%
%
%

%
%
%
%
%
%
%
























































%
%
%
%
%
%

%
%
%
%
%
%











%
%
%
%

%
%
%
%
%
%
%
%
%
%
%
%
%
%
%
%
%
%

%
%
%
%
%
%
%

%
%

%
%
%
%

%
%
%
%
%
%
%
%


%
%
%
%
%
%
%
%
\end{thebibliography}
\end{document}